# High Throughput of WiMAX MIMO-OFDM Including Adaptive Modulation and Coding

Hadj Zerrouki[1], Mohamed Feham[2]
Laboratoire de Systèmes de Technologies de l'Information
et de Communication (STIC)
University Abou Baker Belkaid, Tlemcen, Algeria.

*Abstract*— **WiMAX technology is based on the IEEE 802.16 specification of which IEEE 802.16-2004 and 802.16e amendment are Physical (PHY) layer specifications. IEEE 802.16-2004 currently supports several multiple-antenna options including Space-Time Codes (STC), Multiple-Input Multiple-Output (MIMO) antenna systems and Adaptive Antenna Systems (AAS). The most recent WiMAX standard (802.16e) supports broadband applications to mobile terminals and laptops. Using Adaptive Modulation and Coding (AMC) we analyze the performance of OFDM physical layer in WiMAX based on the simulation results of Bit-Error-Rate (BER), and data throughput. The performance analysis of OFDM-PHY is done. In this paper, an extension to the basic SISO mode, a number of 2x2 MIMO extensions are analysed under different combinations of digital modulation (QPSK, 16-QAM and 64-QAM) and Convolutional Code (CC) with 1/2 , 2⁄3 and 3/4 rated codes. The intent of this paper is to provide an idea of the benefits of multiple antenna systems over single antenna systems in WiMAX type deployments.**

## I. Introduction

The first WiMAX systems were based on the IEEE 802.16-2004 standard [1]. This targeted fixed broadband wireless applications via the installation of Customer Premises Equipment (CPE). In December, 2005 the IEEE completed the 802.16e-2005 [2] amendment, which added new features to support mobile applications. The resulting standard is commonly referred to as mobile WiMAX.

An ever crowded radio spectrum implies that future demands must be met using more data throughput wireless technologies. Since system bandwidth is limited and user demand continues to grow, spectral efficiency is vital. One way to improve link capacity, and potentially increase spectral efficiency, is the application of MIMO. It is well reported in the literature that MIMO physical (PHY) layer techniques have the potential to significantly increase bandwidth efficiency in a rich scattering environment [3]. Orthogonal Frequency Division Multiplexing (OFDM) is a well-established technique for achieving low-cost broadband wireless connectivity, and has been chosen as the air interface for a range of new standards, including IEEE802.16d/e. The ideas of MIMO and OFDM have been combined by a number of authors to form a new class of MIMO-OFDM system [4] [5]. This approach represents a promising candidate for WiMAX applications.

WiMAX standard supports a full-range of smart antenna techniques, including spatial transmit diversity and spatial multiplexing (SM). Spatial transmit diversity is achieved by applying Alamouti's Space-Time coding. SM can also be employed to increase the error-free peak throughput. Higher order modulation schemes with SM increase the link throughput, but require high SNR to achieve low Packet Error Rates (PER). Space-Time Block Coding (STBC) provides strong diversity gain, but cannot increase the link throughput without the use of Adaptive Modulation and Coding (AMC), and therefore AMC has become a standard approach in recently developed wireless standards, including WiMAX. The idea behind AMC is to dynamically adapt the modulation and coding scheme to the channel conditions so as to achieve the highest spectral efficiency at all times. Adaptive modulation changes the coding scheme and/or modulation method depending on channel-state information - choosing it in such a way that it squeezes the most out of what the channel can transmit.

This paper investigates the performance of the WiMAX standard when MIMO techniques are applied. Bit Error Rate (BER) and throughput results are presented for a MIMO with OFDM system that uses the coding and modulation schemes defined in the WiMAX standard IEEE 802.16-2004. Results are compared with basic SISO operation.

## II. The WiMAX PHY Description

The IEEE 802.16 standard was firstly designed to address communications with direct visibility in the frequency band from 10 to 66 GHz. Due to the fact that non-line-of-sight transmissions are difficult when communicating at high frequencies, the amendment 802.16a was specified for working in a lower frequency band, between 2 and 11 GHz. The IEEE 802.16d specification is a variation of the fixed standard (IEEE 802.16a) with the main advantage of optimizing the power consumption of the mobile devices. The last revision of this specification is better known as IEEE 802.16-2004 [1].

On the other hand, the IEEE 802.16e standard is an amendment to the 802.16-2004 base specification with the aim of targeting the mobile market by adding portability.

WiMAX standard-based products are designed to work not only with IEEE 802.16-2004 but also with the IEEE 802.16e specification. While the 802.16-2004 is primarily intended for stationary transmission, the 802.16e is oriented to both stationary and mobile deployments.

### A. *PHY Layer Overview*

WiMAX is not truly new; rather, it is unique because it was designed from the ground up to deliver maximum throughput to maximum distance while offering 99.999 percent reliability. To achieve this, the designers (IEEE 802.16 Working Group D) relied on proven technologies for the PHY including orthogonal frequency division multiplexing (OFDM), time division duplex (TDD), frequency division duplex (FDD), Quadrature Phase Shift Keying (QPSK), and Quadrature Amplitude Modulation (QAM), to name only a few. WiMAX has a scalable physical-layer architecture that allows for the data rate to scale easily with available channel bandwidth. This scalability is supported in the OFDMA[1] mode, where the FFT (fast Fourier transform) size may be scaled based on the available channel bandwidth. For example, a WiMAX system may use 128, 512, or 1024 FFTs based on whether the channel bandwidth is 1.25MHz, 5MHz, or 10MHz respectively. This scaling may be done dynamically to support user roaming across different networks that may have different bandwidth allocations.

### B. *OFDM Parameters in WiMAX*

As mentioned previously, the fixed and mobile versions of WiMAX have slightly different implementations of the OFDM physical layer. Fixed WiMAX, which is based on IEEE 802.16-2004, uses a 256 FFT-based OFDM physical layer. Mobile WiMAX, which is based on the IEEE 802.16e-2005[2] standard, uses a scalable OFDMA-based physical layer. In the case of mobile WiMAX, the FFT sizes can vary from 128 bits to 2048 bits.

Table I shows the OFDM-related parameters for both the OFDM-PHY and the OFDMA-PHY. The parameters are shown here for only a limited set of profiles that are likely to be deployed and do not constitute an exhaustive set of possible values.

### C. *WiMAX OFDMA-PHY*

In Mobile WiMAX, the FFT size is scalable from 128 to 2048. Here, when the available bandwidth increases, the FFT size is also increased such that the subcarrier spacing is always 10.94 kHz. This keeps the OFDM symbol duration, which is the basic resource unit, fixed and therefore makes scaling have minimal impact on higher layers. A scalable design also keeps the costs low. The subcarrier spacing of 10.94 kHz was chosen as a good balance between satisfying the delay spread and Doppler spread requirements for operating in mixed fixed and mobile environments. This subcarrier spacing can support delay-spread values up to 20 µs and vehicular mobility up to 125 km/h when operating in 3.5GHz. A subcarrier spacing of 10.94 kHz implies that 128, 512, 1024, and 2048 FFT are used when the channel bandwidth is 1.25MHz, 5MHz, 10MHz, and 20MHz, respectively. It should, however, be noted that mobile WiMAX may also include additional bandwidth profiles.

TABLE I. OFDM PARAMETERS USED IN WIMAX

| Parameter | Fixed WiMAX OFDM | Mobile WiMAX Scalable OFDMA[a] | | | |
|---|---|---|---|---|---|
| FFT size | 256 | 128 | **512** | 1024 | 2048 |
| Number of used data subcarriers | 192 | 72 | **360** | 720 | 1440 |
| Number of pilot subcarriers | 8 | 12 | **60** | 120 | 240 |
| Number of null/guardband subcarriers | 56 | 44 | **92** | 184 | 360 |
| Cyclic prefix or guard time | 1/4, **1/8**, 1/16, 1/32 | | | | |
| Channel bandwidth (MHz) | 3.5 | 1.25 | **5** | 10 | 20 |
| Subcarrier frequency spacing (kHz) | 15.625 | **10.94** | | | |
| Useful symbol time (µs) | 64 | **91.4** | | | |
| Guard time assuming 12.5% (µs) | 8 | **11.4** | | | |
| OFDM symbol duration (µs) | 72 | **102.9** | | | |
| Number of OFDM symbols in 5 ms frame | 69 | **48.0** | | | |

a. Boldfaced values correspond to the OFDMA parameters used in our evaluation of the **Fixed WiMAX** standard.

## III. SYSTEM MODEL DESCRIPTION

Since increase the link throughput systems is the main goal of this work. Figure 1 depicts the block scheme of a typical structure of a system combining Space Time Coding (STC) with MIMO OFDM enabled WiMAX simulator used in this paper. The Block diagram represents the whole system model or the signal chain at base band. The block system is divided into 2 main sections namely the transmitter and the receiver.

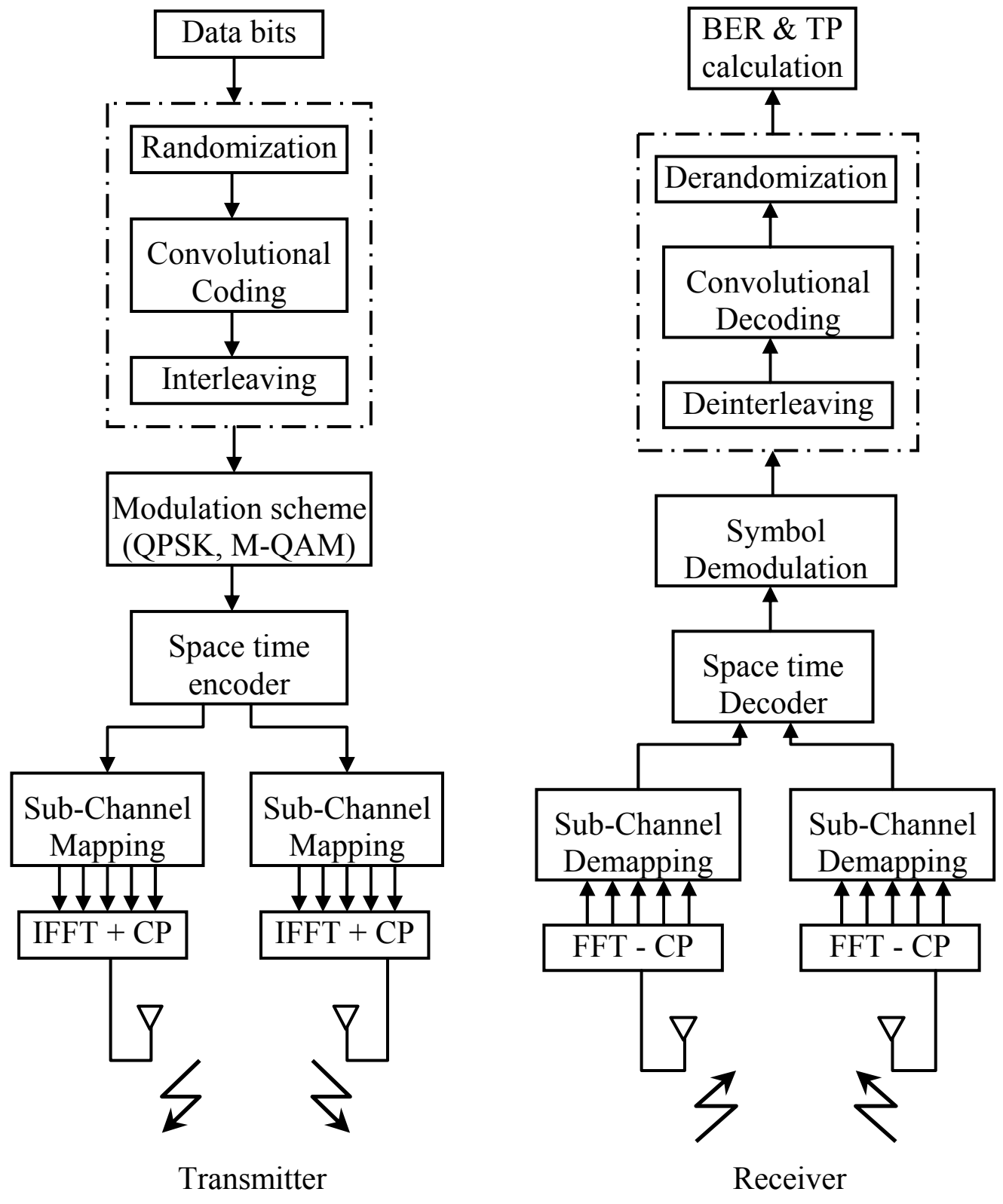

Figure 1. System block diagram for WiMAX MIMO OFDM simulator used in this paper.

[1] Orthogonal Frequency Division Multiplexing Access.
[2] Although the scalable OFDMA scheme is referred to as mobile WiMAX, it can be used in fixed, nomadic, and mobile applications.

### A. Transmitter

The data is generated from a random source, consists of a series of ones and zeros. Since the transmission is done block wise, when forward error correction (FEC) is used, the size of the data generated depends on the block size used, modulation scheme used to map the bits to symbols (QPSK, *M*-QAM), and whether FEC is used or not [1]. The generated data is passed on to the next stage, either to the FEC block or directly to the symbol mapping if FEC is not used.

*1) Channel coding:* There are various combinations of modulations and code rates available in the OFDMA burst. Channel coding includes the randomization of data, forward error correction (FEC) encoding, interleaving, and modulation. In some cases, transmitted data may also be repeated on an adjacent subcarrier.

*2) Randomization:* Randomization of the data sequence is typically implemented to avoid the peak-to-average power ratio (PAPR) increasing beyond that of Gaussian noise, thus putting a boundary on the nonlinear distortion created in the transmitter's power amplifiers. It can also help minimize peaks in the spectral response.

*3) Forward error correction (FEC):* In our case, the error correcting codes are used, the data generated is randomized so as to avoid long run of zeros or ones, the result is ease in carrier recovery at the receiver. The randomized data is encoded using tail biting convolutional codes (CC) whose constraint length is 7 and the native code rate is ½ (puncturing of codes is provided in the standard to produce higher code rates). Finally interleaving is done by two stage permutation, first to avoid mapping of adjacent coded bits on adjacent subcarriers and the second permutation insures that adjacent coded bits are mapped alternately onto less or more significant bits of the constellation, thus avoiding long runs of lowly reliable bits.

*4) Modulation:* There are three modulation types available for modulating the data onto the subcarriers : QPSK, 16QAM, and 64QAM used with gray coding in the constellation map. In the Uplink, the transmit power is automatically adjusted when the modulation coding sequence (MCS) changes to maintain the required nominal carrier-to-noise ratio at the BS receiver. 64QAM is not mandatory for the Uplink.

*5) Space Time Encoder (MIMO encoder):* The Space Time Encoder stage converts one single input data stream into multiple output data streams. How the output streams are formatted depends on the type of MIMO method employed. Different symbols are simultaneously transmitted over these antennas to reduce noise interference. The receiver after receiving the signal retrieves the bits using Maximum Likelihood decoding algorithm and passes the data to the guard band removal block.

*6) IFFT and cyclic prefixand:* An'*N*' point inverse discrete fourier transform (IDFT) of '*X(k)*'is defined as:

$$x(n) = \frac{1}{N}\sum_{n=0}^{N-1} x(k) e^{j\frac{2\pi kn}{N}} \quad \text{for } `n' = 0,1,\ldots N\text{-}1. \tag{1}$$

From the equation we can infer that this is equivalent to generation of OFDM symbol. An efficient way of implementing IDFT is by inverse fast Fourier transform (IFFT). Hence IFFT is used in generation of OFDM symbol. The addition of cyclic prefix is done on the time domain symbol obtained after IFFT. The IFFT size ('*N*' value) is considered as 512 in simulations. This data is fed to the channel which represents 'Rayleigh fading channel model' and also implements multipath as shown in block diagram.

### B. Receiver

The first thing done at receiver (in simulation) is removal of cyclic prefix, thus eliminating the inter symbol interference (ISI). Data is then passed through the serial to parallel converter of size 512 and then fed to the FFT for frequency domain transformation. The signal was distorted by the channel, to reconstruct the original signal we need information as to how the channel acted on the transmitted signal so that we can mitigate its effect. This is called equalization. In an OFDM system, this is done by channel estimation and interpolation, and reverse process (including deinterleaving and decoding) is executed to obtain the original data bits. As the deinterleaving process only changes the order of received data, the error probability is intact.

### C. Rayleigh Fading Channel

Rayleigh Fading is one kind of statistical model which propagates the environment of radio signal. According to Rayleigh distribution magnitude of a signal which has passed though the communication channel and varies randomly. Rayleigh Fading works as a reasonable model when many objects in environment which scatter radio signal before arriving of receiver. When there is no propagation dominant during line of sight between transmitter and receiver on that time Rayleigh Fading is most applicable. On the other hand Rician Fading is more applicable than Rayleigh Fading when there is dominant line of sight. During our simulation we used Rayleigh Fading when we simulate the performance of BER and throughput Vs Signal to Noise Ratio.

## IV. Enhancement with MIMO link adaptation

It is well-understood that spectral efficiency is the key to good system design. Without loss of generality, the normalized effective system efficiency can be written as:

$$\xi_{sys} = \frac{\textit{system data throughput}}{\textit{total radio resouces allocated}} \tag{2}$$

Two clear approaches emerge to improve the effective efficiency. Firstly the system data throughput can be improved by using methods such as high level AMC and MIMO. Secondly, improvements can be made to reduce the amount of radio resource required in the system. This section focuses on the system performance enhancements with MIMO Link Adaptation (LA) in combination with OFDMA. Using a statistical approach, it is possible to demonstrate the potential benefits of a relay enhanced mobile WiMAX deployment.

### A. *Throughputs and coverages for WiMAX system*

It is well-known that MIMO promises transmission efficiency enhancement which achieves one aspect of efficiency enhancement. IEEE 802.16e standard supports a full range of smart antenna technologies. Together with modulation and coding, the link throughput for each user can be calculated from Packet Error Rate (*PER*) by:

$$C_{link} = \frac{N_D N_b R_{FEC} R_{STC}}{T_s} \times (1 - PER) \quad (3)$$

Where, $T_s$, $N_D$, $N_b$, $R_{FEC}$ and $R_{STC}$ denote the OFDMA symbol duration, the number of assigned data subcarriers, the number of bits per subcarrier, FEC coding rate, and space-time coding rate for the user. Equation (3) implies that, a combination of MIMO, AMC and flexible sub-channelization is required to maximize the link performance.

### B. *MIMO scenarios description*

*1) Space-Time Block Coding (STBC):* Our Fixed WIMAX simulator implements the Alamouti scheme [7] on the Downlink to provide transmit and receive diversity. This scheme uses a transmission matrix [$s_1$, $-s_2^*$; $s_2$, $s_1^*$], where $s_1$ and $s_2$ represents two consecutive OFDMA symbols.

*2) Spatial Multiplexing (SM):* WiMAX system supports SM to increase the peak error free data rate [8]. The idea behind spatial multiplexing is that multiple independent streams can be transmitted in parallel over multiple antennas and can be separated at the receiver using multiple receive chains through appropriate signal processing. This can be done as long as the multipath channels as seen by the various antennas are sufficiently decorrelated, as would be the case in a scattering-rich environment. Spatial multiplexing provides data rate and capacity gains proportional to the number of antennas used, a 2x2 SM system can double the peak data rate. This comes at the expense of sacrificing diversity gain, and hence a much higher SNR is required.

*3) Adaptive Modulation and Coding:* WiMAX systems use adaptive modulation and coding in order to take advantage of fluctuations in the channel. The basic idea is quite simple: Transmit as high a data rate as possible when the channel is good, and transmit at a lower rate when the channel is poor, in order to avoid excessive dropped packets. Lower data rates are achieved by using a small constellation, such as QPSK, and low-rate error-correcting codes, such as rate convolutional or turbo codes. The higher data rates are achieved with large constellations, such as 64 QAM, and less robust error correcting codes; for example, rate convolutional, turbo, or LDPC codes. In all, 52 configurations of modulation order and coding types and rates are possible, although most implementations of WiMAX offer only a fraction of these[9].

In our case, by using six (6) of the common WiMAX burst profiles; it is possible to achieve a large range of spectral efficiencies. This allows the throughput to increase as the signal-to-interference-plus-noise ratio (SINR) increases following the trend promised by Shannon's formula. In this case, the lowest offered data rate is QPSK and rate 1/2 Convolutional codes; the highest data-rate burst profile is with 64 QAM and rate 3/4 Convolutional codes. The achieved throughput normalized by the bandwidth is defined as:

$$T = (1 - BLER)_r \log_2(M) \quad bps/Hz \quad (4)$$

where *BLER* is the block error rate, $r \leq 1$ is the coding rate, and *M* is the number of points in the constellation. For example, 64 QAM with rate 3/4 codes achieves a maximum throughput, when BLER→0; QPSK with rate 1/2 codes achieves a best-case throughput.

## V. SIMULATION RESULTS

In this section SISO and MIMO BER and Throughput results are presented using the Fixed WiMAX simulator. The Simulation model was implemented in Matlab® 7. The PHY parameters used in simulation are given in Table I, with boldfaced values correspond to the OFDMA-PHY parameters. A carrier frequency of 2GHz is considered. For Spatial Multiplexing, an MMSE receiver is used to remove the inter-stream interference on a per sub-carrier basis. The link throughput is calculated from the PER as given by equation (3) in section IV. On the downlink, we consider that no sharing of OFDMA symbol. In this case of single-user MIMO, the multiple streams are intended for the same receiver; we have considered transmission formats with only a single stream for a single user as a basic OFDM in fixed WiMAX.

Figure 2 and 3 displays the performance of SISO and 2x2 MIMO STBC systems. For comparison purposes, the SNR versus BER graph is plotted for different modulation schemas and Convolutional Coding (CC) rates are showed.

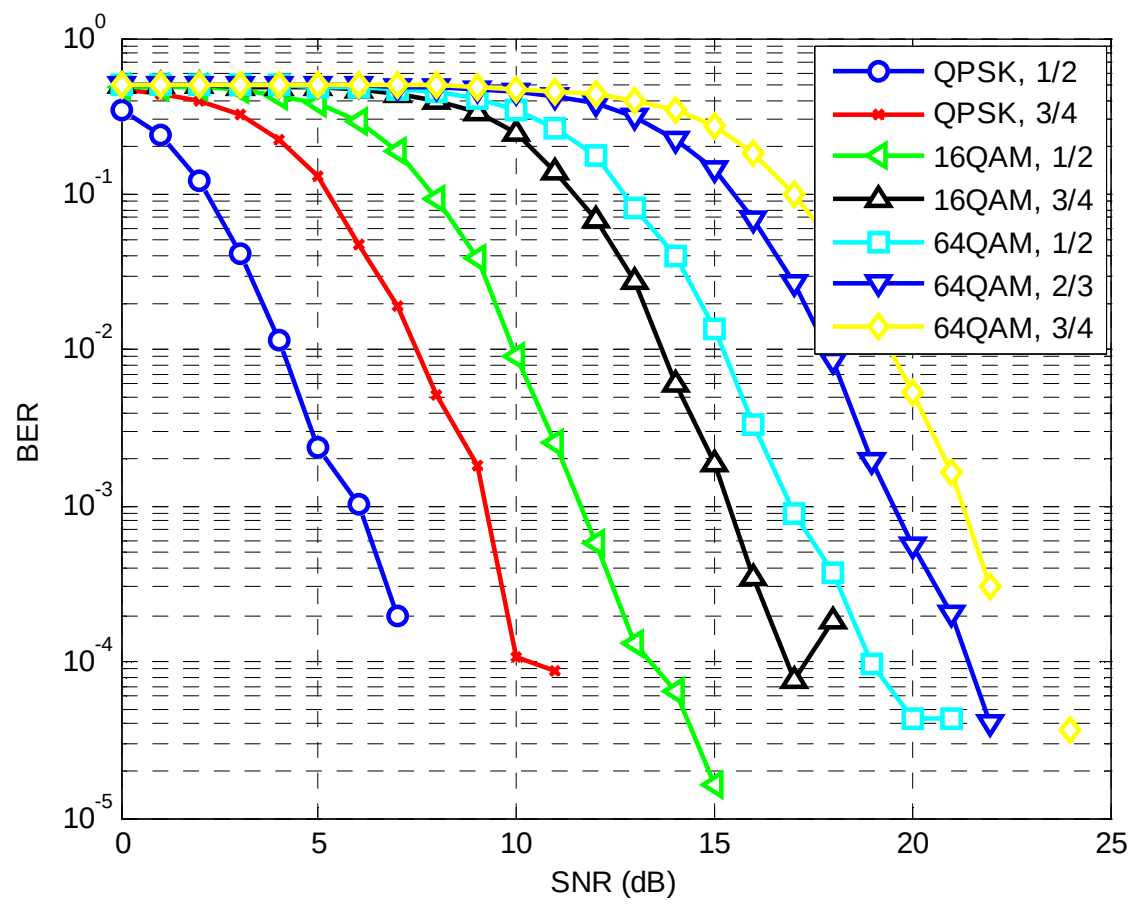

Figure 2. BER performance of SISO WiMAX system.

It is evident from the graph that the QPSK modulation scheme with 1/2 CC rate is better suited for OFDM transmission in terms of BER performance, and it can be see that the BER performance MIMO system with 2x2 STBC yields a gain of about 3 dB over the corresponding SISO system at a BER of $10^{-3}$.

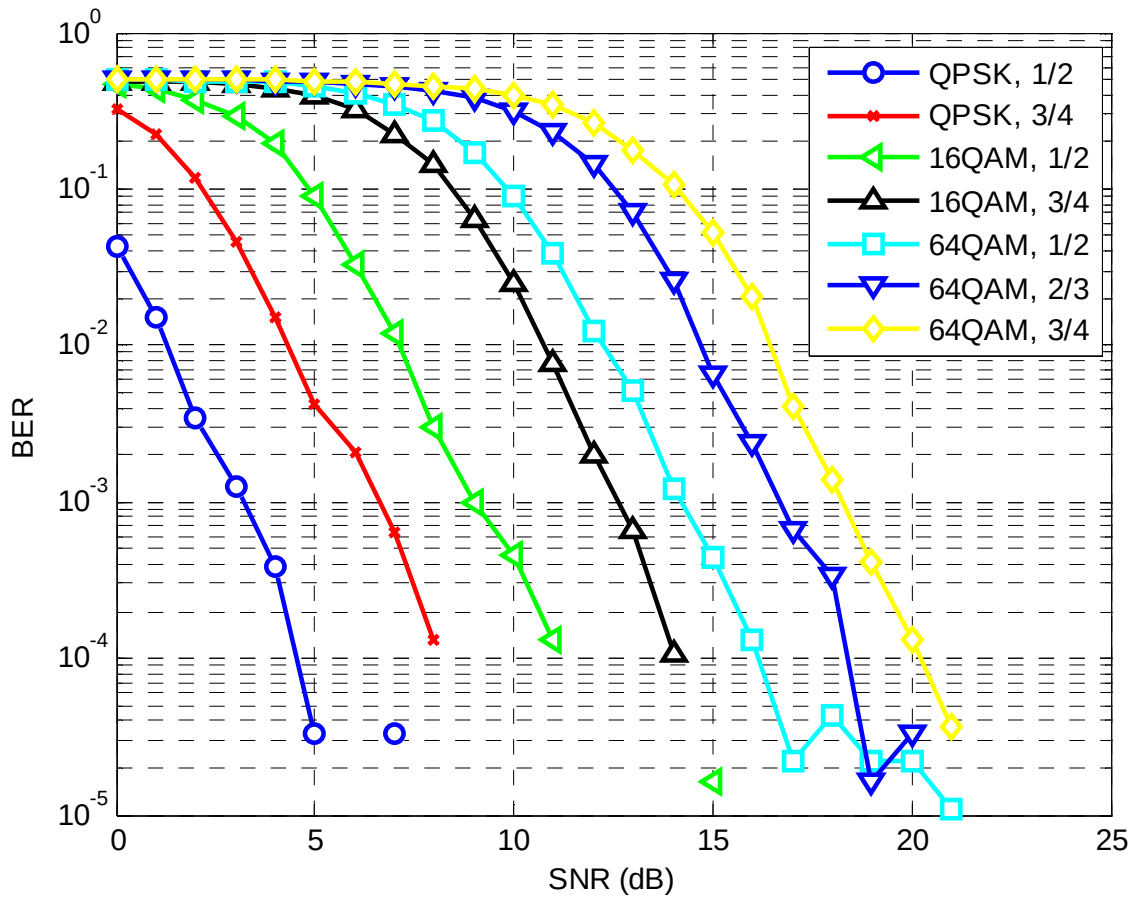

Figure 3. BER performance of 2x2 STBC WiMAX system.

Figure 4 and 5 present the throughput versus SNR graphs for the SISO and MIMO 2x2 STBC scenarios respectively. We observe that STBC offers a significant performance gain of 3dB to 4dB, the exact value depend on the selected link-speed. As we see here, STBC does not improve the data throughput, however at a given SNR STBC can provide a significant increase in throughput when combined with suitable link adaptation, since higher throughput modes can be used at much lower values of SNR. STBC increases the robustness of the WiMAX system by coding over the different transmitter branches and over space and temporal dimension. In this way, the spectral efficiency of MIMO is exploited by adding extra redundancy to improve the performance.

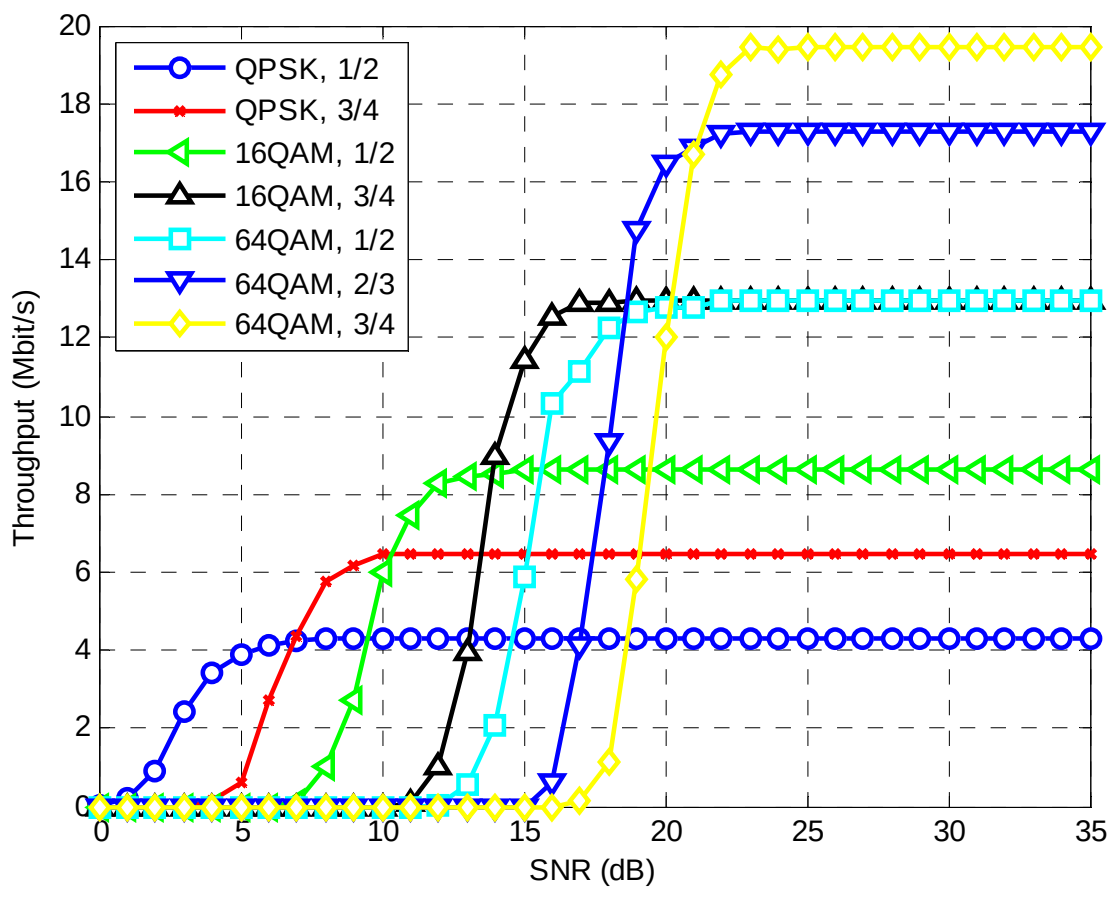

Figure 4. Throughput of SISO WiMAX system.

Figure 6 illustrates the simulated MIMO 2x2 SM system throughputs versus SNR. The main advantage of SM is that it directly exploits the MIMO channel capacity to improve the data throughput by simultaneously transmit different signals on different transmit antennas, at the same carrier frequency. The main disadvantage is that no redundancy is added and, thus, it might suffer from poor link reliability. To overcome this problem additional channel coding can introduced. This, however, reduces its data rate advantage. As expected, the SM 2x2 modes doubles the peak error-free throughput of every link-speed. However, at low SNR values the throughput of SM is less than STBC.

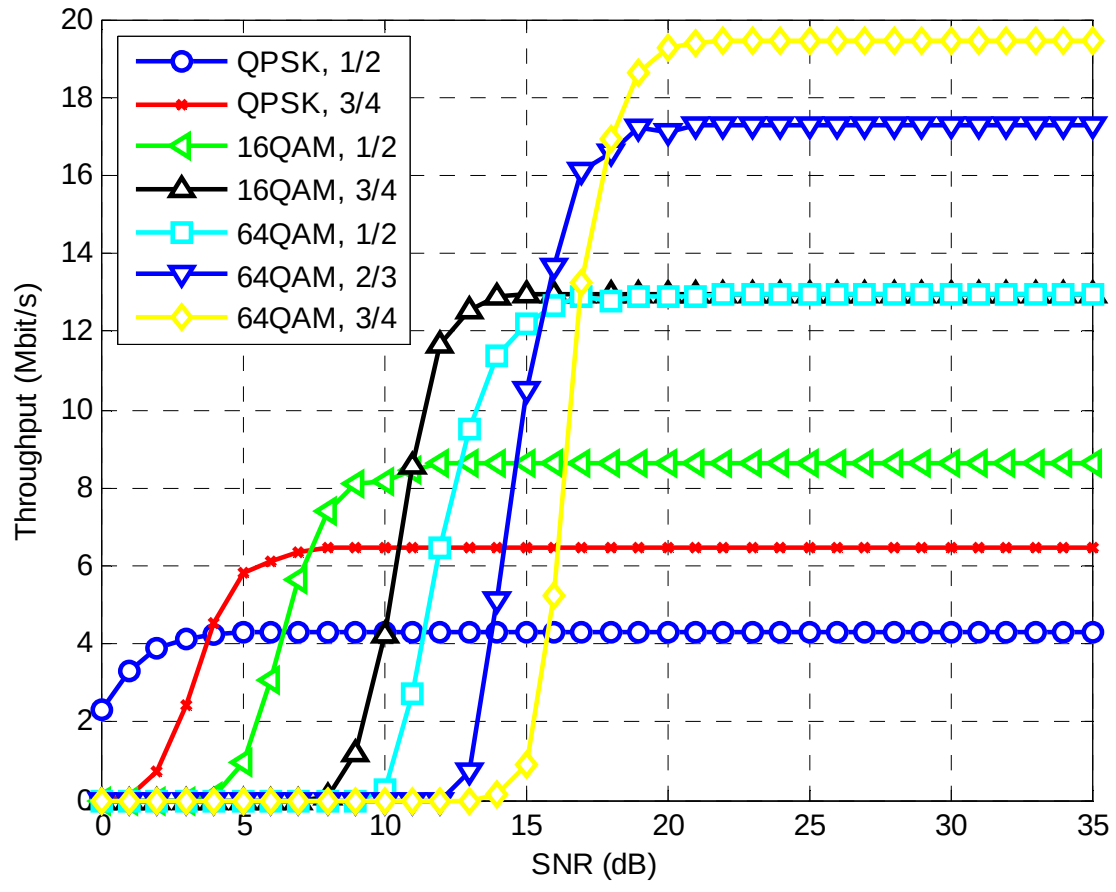

Figure 5. Throughput of 2x2 STBC WiMAX system.

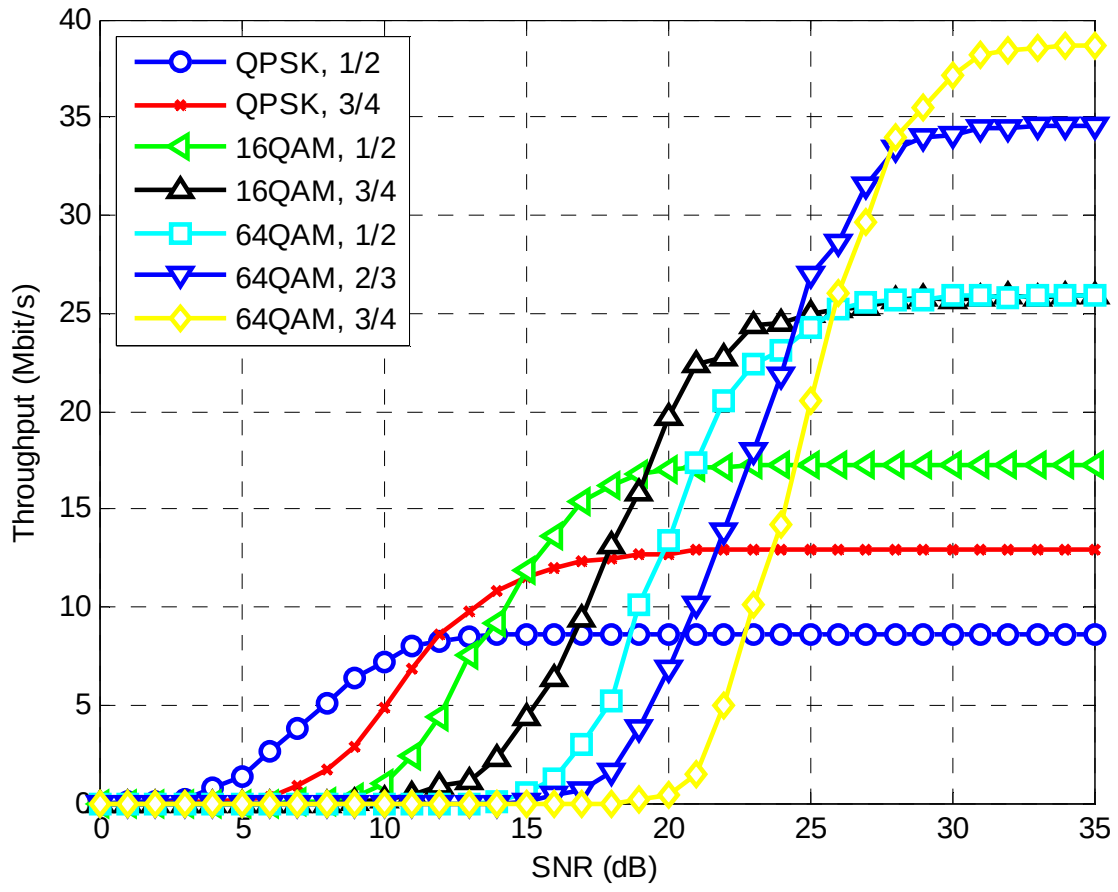

Figure 6. Throughput of 2x2 SM WiMAX system.

Figure 7 shows the throughput envelope versus SNR for all the investigated fixed WiMAX scenarios: SISO, 2x2 STBC, and 2x2 SM. This envelope assumes the use of adaptive modulation and coding (AMC) to maximize the expected throughput. Obviously, both MIMO schemes outperform the SISO scenario. However, for a very spatially correlated channel, the SM method can be worse than SISO. In this case STBC performance would tend to that of SISO. The STBC produces the best performance at low to medium values of SNR, due to its robustness in poor channel conditions. On the other hand, at high SNR the increased error-free data rate makes SM the best choice.

WiMAX system supports Adaptive MIMO Switching (AMS) to select the best MIMO scheme. Figure 7 clearly shows that for the channel conditions analysed here, the switching point between STBC and SM is 20dB. This value will increase with increasing spatial correlation.

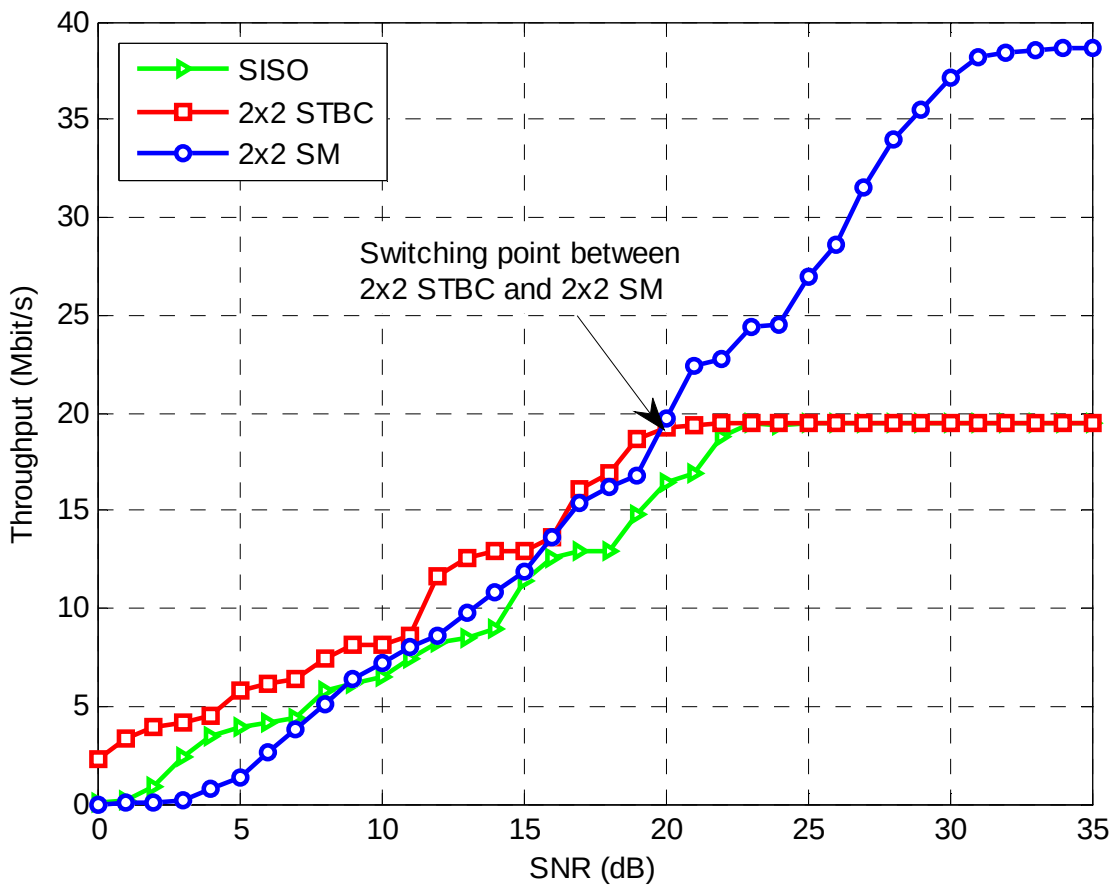

Figure 7. Adaptive Modulation and Coding (AMC) and switching point between STBC 2x2 and SM 2x2.

## VI. CONCLUSION

In this paper, we considered the calculation of data throughput of the downlink of OFDMA-based IEEE802.16 WiMAX systems; this paper has presented a detailed study of the benefit of MIMO when applied to WiMAX. Throughput results were presented for diverse scenarios. At lower values of SNR STBC is preferred. However, at high SNR Adaptive MIMO Switching should be used to switch to SM. Give that SM 2x2 doubles the error-free throughput, at high SNR this scheme leads to the highest throughput. In practice, the viability of SM (and the value of the SNR switching threshold) depends on the level of spatial correlation.